\documentclass[prd,twocolumn,a4paper,superscriptaddress,floatfix]{revtex4}
\usepackage{graphicx,amssymb}
\usepackage[usenames]{color}
\begin{document}
\color{red}

\newcommand{\bb}{\bibitem}
\newcommand{\ov}{\overline}
\newcommand{\wt}{\widetilde}
\newcommand{\nn}{\nonumber\\}
\newcommand{\bfi}{\begin{figure}}
\newcommand{\efi}{\end{figure}}
\newcommand{\sech}{\mbox{sech}}
\newcommand{\arcsinh}{\mbox{arcsinh}}
\newcommand{\arccosh}{\mbox{arccosh}}

\newcommand{\be}{\begin{equation}}
\newcommand{\ee}{\end{equation}}
\newcommand{\bq}{\begin{eqnarray}}
\newcommand{\eq}{\end{eqnarray}}
\newcommand{\bsq}{\begin{subequations}}
\newcommand{\esq}{\end{subequations}}
\newcommand{\bc}{\begin{center}}
\newcommand{\ec}{\end{center}}
\newcommand {\R}{{\mathcal R}}
\newcommand{\al}{\alpha}
\newcommand\lsim{\mathrel{\rlap{\lower4pt\hbox{\hskip1pt$\sim$}}
    \raise1pt\hbox{$<$}}}
\newcommand\gsim{\mathrel{\rlap{\lower4pt\hbox{\hskip1pt$\sim$}}
    \raise1pt\hbox{$>$}}}

\title{Interfaces with internal structures in generalized rock-paper-scissors models}
\author{P.P. Avelino}
\affiliation{Centro de Astrof\'{\i}sica da Universidade do Porto, 4150-762 Porto, Portugal}
\affiliation{Departamento de F\'{\i}sica e Astronomia, Faculdade de Ci\^encias, Universidade do Porto, 4169-007 Porto, Portugal}
\author{D. Bazeia}
\affiliation{Departamento de F\'{\i}sica, Universidade Federal da Para\'{\i}ba 58051-970 Jo\~ao Pessoa, PB, Brazil}
\author{L. Losano}
\affiliation{Departamento de F\'{\i}sica, Universidade Federal da Para\'{\i}ba 58051-970 Jo\~ao Pessoa, PB, Brazil}
\author{J. Menezes}  
\affiliation{Escola de Ci\^encias e Tecnologia, Universidade Federal do Rio Grande do Norte\\
Caixa Postal 1524, 59072-970, Natal, RN, Brazil}
\affiliation{Institute for Biodiversity and Ecosystem Dynamics, University of Amsterdam, Science Park 904, 1098 XH Amsterdam, The Netherlands}
\author{B.F. de Oliveira}  
\affiliation{Departamento de F\'{\i}sica, Universidade Estadual de Maring\'a, 87020-900 Maring\'a, PR, Brazil}


\date{\today}
\begin{abstract}
In this work we investigate the development of stable dynamical structures along interfaces separating domains belonging to enemy partnerships, in the context of cyclic predator-prey models with an even number of species $N \ge 8$. We use both stochastic and field theory simulations in one and two spatial dimensions, as well as analytical arguments, to describe the association at the interfaces of mutually neutral individuals belonging to enemy partnerships and to probe their role in the development of the dynamical structures at the interfaces. We identify an interesting behaviour associated to the symmetric or asymmetric evolution of the interface profiles depending on whether $N/2$ is odd or even, respectively. We also show that the macroscopic evolution of the interface network is not very sensitive internal structure of the interfaces. Although this work focus on cyclic predator prey-models with an even number of species, we argue that the results are expected to be quite generic in the context of spatial stochastic May-Leonard models.
\end{abstract}

\maketitle

\section{Introduction}

The development of diversity in nature results in multi-scale dynamics associated to cooperation, mobility and competition between a large number of species in many different scenarios; see e.g., Refs.~\cite{smith82,nowak06evolutionaryDynamicsBOOK,sole2006selforganization,Szabó200797}. The development of macroscopic complexity seems to spring very naturally even in the case of very simple cyclic predator-prey models with a low number of species, as in the case of the classic rock-paper-scissor (RPS) game \cite{PhysRevLett.77.2125,Kerr2002,Kirkup2004,Reichenbach2007,PhysRevLett.99.238105}. The RPS model describes the evolution of three species in cyclic interaction and if the population mobility is small enough, the spatial RPS model has been shown to allow for the stable coexistence of the three species with the formation of complex patterns \cite{Kerr2002,Kirkup2004,Reichenbach2007,PhysRevLett.99.238105}. See also Refs.~\cite{May-Leonard,Boerlijst199117,Boerlijst2010,PhysRevE.76.051921,Peltomaki2008,PhysRevE.76.051921,PhysRevE.81.046113,PhysRevE.83.011917,PhysRevE.85.051903,1742-5468-2012-07-P07014,PhysRevE.86.036112,PhysRevE.86.031119,PhysRevE.87.032148,Avelino2014393,0305-4470-38-30-005,PhysRevE.64.042902,Lutz2013286} for other investigations of direct interest to the current work.

The basic interactions behind the RPS model are motion, reproduction, and predation, but generalisations incorporating
new interactions and further species have also been proposed in the literature \cite{PhysRevE.76.051921,Peltomaki2008,PhysRevE.85.051903,PhysRevE.86.036112,PhysRevE.86.031119,PhysRevE.87.032148,Szabo2008,1742-5468-2012-07-P07014}. We learned from these investigations that the increase of the number of species generally leads to the development of more complex dynamical patterns. In particular, in \cite{PhysRevE.86.036112,PhysRevE.86.031119} it has been shown that the spatial structure and dynamics of population networks is extremely dependent both on the predator-prey interaction rules (leading in many cases to the development of partnerships between individuals of different species), and on the number of competing species. These studies inspired Roman, Dasgupta, and Pleimling \cite{PhysRevE.87.032148} to investigate similar models, focusing on the interplay between competition and partnership in spatial environments occupied by a large number of species. They worked to quantify coarsening behaviour and pattern formation, noting the presence of partnerships among distinct species following the maxim that {\it the enemy of my enemy is my friend.} Another interesting effect may appear in predator-prey models defined in three spatial dimensions: the generation of string networks, as recently investigated in \cite{Avelino2014393}.

Here, we study the development of peaceful associations between individuals belonging to enemy partnerships and its effect on the development of dynamical structures along the interfaces separating competing domains. We extend previous work by Szab\'o, Szolnoki and Sznaider \cite{PhysRevE.76.051921} which also noted the development of dynamical structures at the interfaces. Their model, however, did not included the presence of empty sites which is an essential ingredient for the development of the associations studied in the present paper. Other investigations on the dynamics of interfaces in a biological framework were developed in \cite{1742-5468-2012-07-P07014} and \cite{PhysRevE.87.032148} in the case of cyclic predator-prey models with $4$ and $6$ species, respectively.

This paper is organised as follows. In Sec.~II we investigate the development of dynamical structures along interfaces separating enemy partnerships in cyclic predator-prey models using two-dimensional stochastic network simulations. In Sec.~III the results of the previous section are confirmed using mean field theory simulations. In Sec.~IV we investigate in more detail the stability of the dynamical structures at the interfaces, using a combination of one-dimensional mean field theory simulations and analytical arguments. In Sec.~V we focus on the macroscopic evolution of interface networks and determine whether or not it can be affected by the presence of dynamical structures along the interfaces. Finally, we conclude in Sec.~VI.

\section{Stochastic network simulations}

We start by considering a family of spatial stochastic May-Leonard models. In this family, individuals of $N$ species and some empty sites (E) are initially distributed on a square lattice with ${\mathcal N}$ sites. The different species are labeled by $i, j= 1, ...,N$, with the cyclic identification $i= i+ k\, N$ where $k$ is an integer. The number of individuals of the species $i$ ($I_i$) and the number of empty sites ($I_E$) obey the relation $I_E+\sum_{i=1}^{N} I_i={\mathcal N}$. The possible interactions are classified as Motion, Reproduction or Predation, represented by  $ i\ \odot \to \odot\ i\,, $ $ i\ \otimes \to ii\,, $ or $ i\  (i+1) \to i\ \otimes\,, $ respectively, where $\otimes$ represents an empty site and $\odot$ represents an arbitrary individual (of any of the $N$ species) or an empty site. For simplicity, we shall assume that Motion ($m$), Predation ($p$) and Reproduction ($r$) interaction probabilities are the same for all species. A random individual (active) is selected to interact with one of its four nearest neighbours (passive) at each time step. The unit of time $\Delta t=1$ is defined as the time necessary for ${\mathcal N}$ interactions to occur (one generation time).

\begin{figure}
\centering
\includegraphics[scale= 0.9]{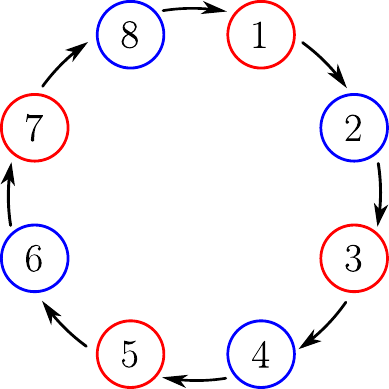}
\caption{(Colour online) Illustration of the cyclic predator-prey rule in the case with $8$ species.} \label{fig1}
\end{figure}

Let us start by focusing on models with an even number of species $N$, following a cyclic predator-prey rule. Unlike the odd $N$ case, which gives rise to spiral patters, even $N$ models produce interface networks without junctions, separating domains with $2$ different partnerships (see, for example, \cite{PhysRevE.86.036112,PhysRevE.86.031119} and references there in). In these models each individual chases and is hunted by only one different species. The predator-prey interactions are illustrated in Fig. \ref{fig1} in the case of a model with $8$ species. 

In this paper we present the results of a large number of network simulations assuming periodic boundary conditions. At the initial time, the number density, $n_i= I_i/{\mathcal N}$, is assumed to be the same for all species, while the number density of empty sites is set to zero, that is $n_E= I_E/{\mathcal N} = 0$. All the stochastic simulations presented in this work have been obtained with $m=0.50$, $r=0.25$ and $p=0.5$, and the snapshots were taken after $5000$ generations. However, we verified that the same qualitative results also hold for other choices of the parameters $m$, $r$ and $p$.

Soon after the simulations start, individuals separate into two partnerships. The maxim {\it the enemy of my enemy is my friend} plays a role as species of a given partnership peacefully share common regions of space, with the battles with the enemy partnership taking place at the domain boundaries. The competition between individuals of different partnerships creates empty sites along the interfaces separating the various domains.

\begin{figure}
\centering
\includegraphics[scale=0.21]{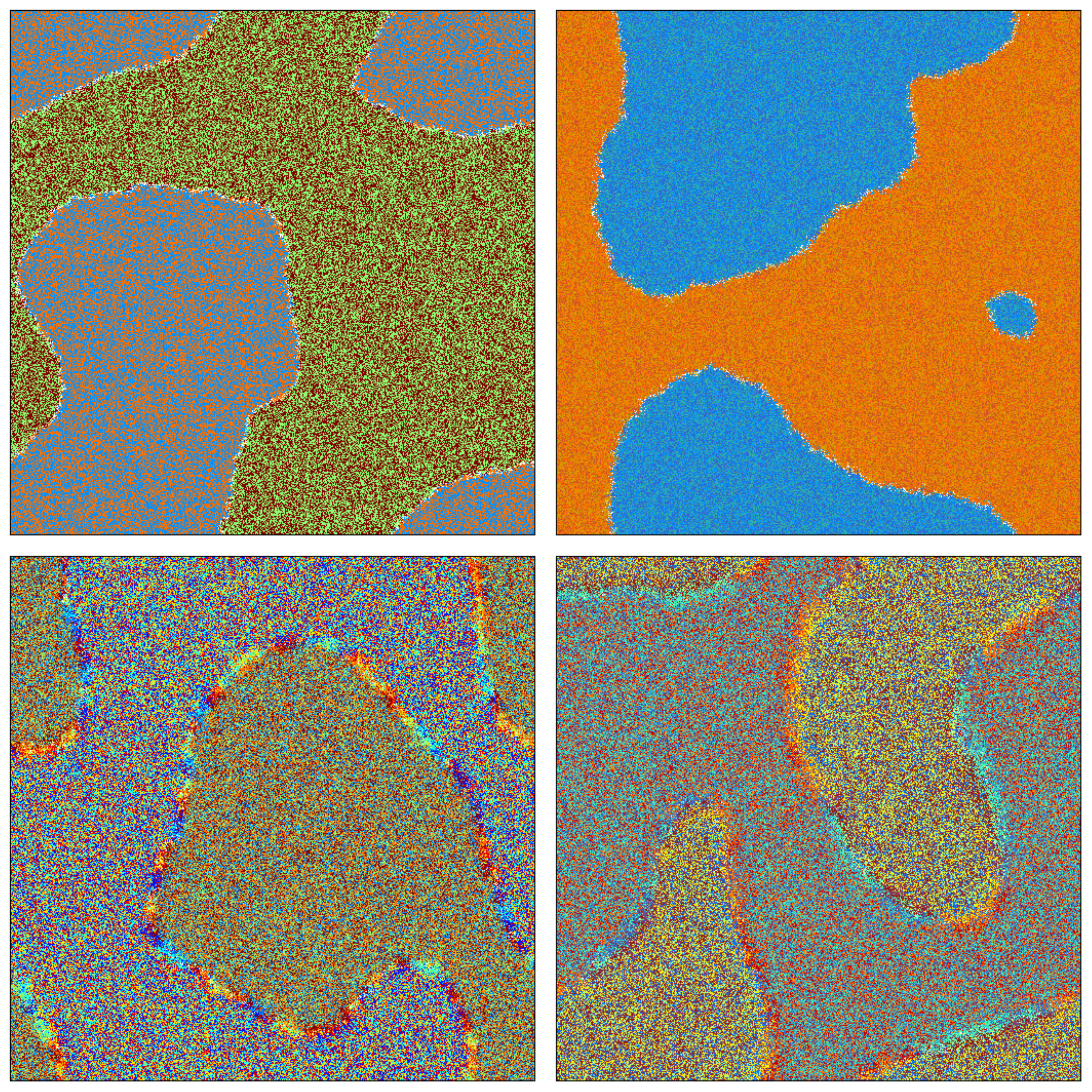}
\caption{(Color online) $512^2$ stochastic network simulations of models with $N=4$ (top left), $N=6$ (top right), $N=8$ (bottom left) and $N=10$ (bottom right). The snapshots were taken after $5000$ generations.}\label{fig2}
\end{figure}

Let us first consider the $N=4$ model, where mutually neutral species aggregate in two partnerships, $\{1,3\}$ and $\{2,4\}$, as shown in the upper left panel of Fig. \ref{fig2}. Note that the distribution of individuals of the two species that aggregate in each partnership is statistically homogeneous inside the respective domain. Although a non-zero mobility  gives rise to intrusions of individuals into enemy domains, the invasion is rapidly put to an end by individuals of the competing domains. For example, individuals of the species $1$ can predate individuals of the species $2$, reproduce and then cross the interface into the enemy domain. They may keep going until they find individuals of the species $4$, ready to defend the territory by killing the invaders. The width of the interface depends on the mobility probability of the species, i.e. the higher the mobility the further the individuals can invade the enemy territory before being caught; consequently, the thicker the interface will be.

We now focus on the $N=6$ model. There are two partnerships, $\{1,3,5\}$ and $\{2,4,6\}$, each occupying separate spatial regions on the lattice. However the species $i$ does not interact with species $i+3$ belonging to the competing partnership. This implies that whenever individuals of the species $i$, present in the battlefront, find individuals of the species $i+3$, they can peacefully share common spatial regions even though they are in a conflict zone and belong to competing partnerships. However, the top right panel in Fig.~\ref{fig2} shows that this is not a stable situation. The frequent attacks of predators from both sides of the interfaces do not allow for long lasting peaceful interactions at the interfaces. Hence, in this case the peaceful associations between species of competing partnerships do not give rise to stable dynamical structures at the interfaces. The top right panel of Fig.~\ref{fig2} shows a snapshot of a simulation of the $N=6$ model. As in the $N=4$ model, the distribution of the species belonging to a given partnership is statistically homogeneous inside the domains, while the individuals fighting at the boundaries give rise to a statistically homogeneous distribution of empty sites along the interfaces.

\begin{figure}
\centering
\includegraphics[scale=0.21]{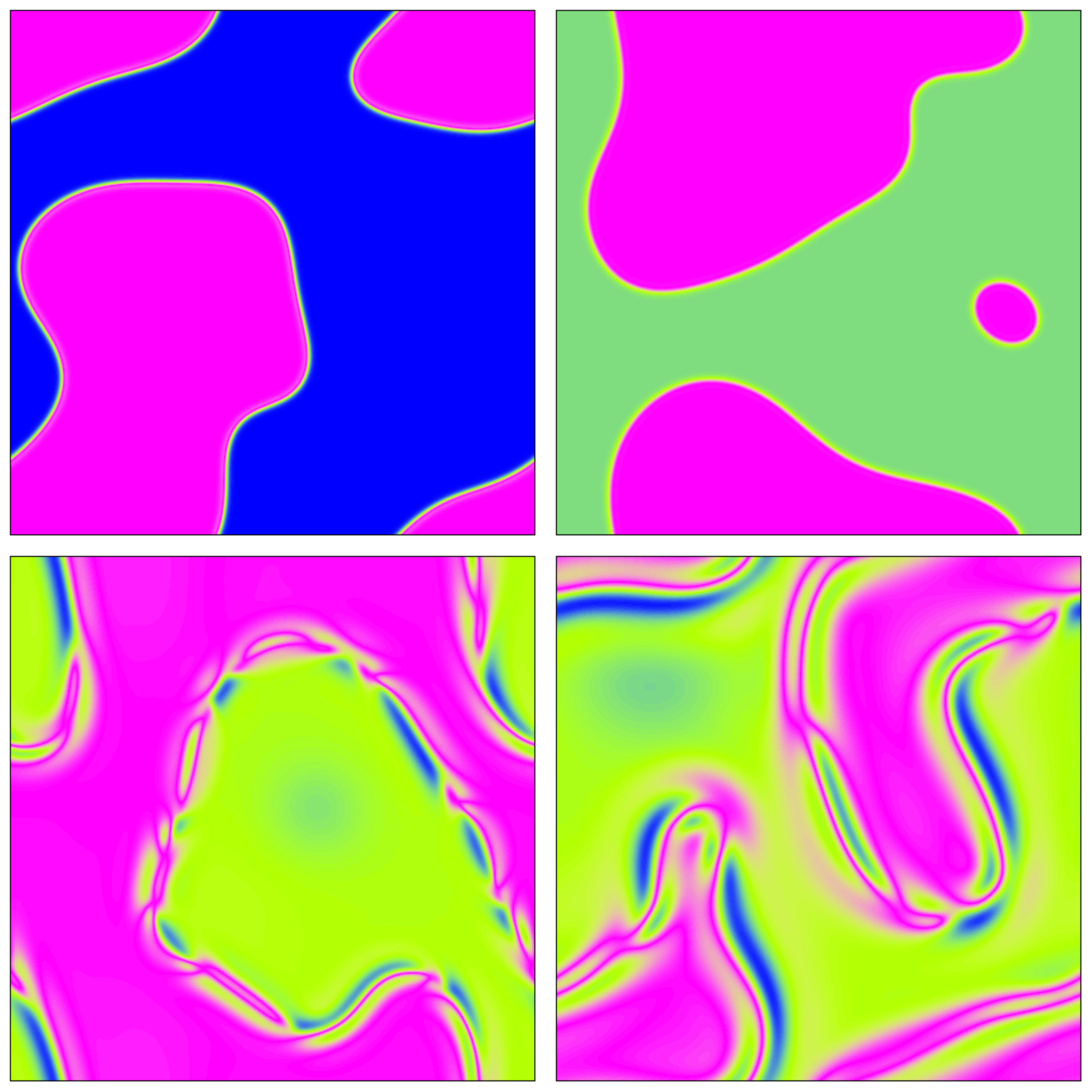}
\caption{(Color online) $512^2$ mean field theory simulations of models with $N=4$ (top left), $N=6$ (top right), $N=8$ (bottom left) and $N=10$ (bottom right). The snapshots were taken after $5000$ generations.}\label{fig3}
\end{figure}

In the $N=8$ model the partnerships $\{1, 3, 5, 7\}$ and $\{2, 4, 6, 8\}$ are formed. In this case there is a larger number of possible peaceful interactions at the interfaces, which leads to the development of stable dynamical structures along the interfaces, as shown in the snapshot at the bottom left panel of Fig. \ref{fig2}. The mixing of colours is always changing (in space and time), as a result of the constant development and destruction, at the interfaces, of the structures made of mutually neutral individuals belonging to competing partnerships. 

In general, an individual of an arbitrary species $i$ can peacefully coexist at the interfaces with species $i \pm k$ belonging to the enemy partnership, where $k$ is an odd integer such that $3\leq k \leq N-5$. In other words, the number of species belonging to the enemy partnership with which the species $i$ can peacefully coexist is $(N-4)/2$. Therefore the larger $N$ is, the more complex will be the behaviour of the dynamical structures formed at the interfaces.

This can be seen in the snapshot obtained for $N=10$ at the bottom left panel of Fig. \ref{fig2}. The larger number of peaceful associations between individuals of species belonging to enemy partnerships ($\{1,3,5,7,9\}$ and $\{2,4,6,8,10\}$) compared to the $N=8$ case, results in more complex interface profiles which we will investigate in more detail in the forthcoming sections.


\section{Mean field theory simulations}


Let us now investigate cyclic predator-prey models using mean field theory simulations. Consider $N+1$ scalar fields ($\phi_0$, $\phi_1$, $\phi_2$, $\ldots$,$\phi_N$) representing the fraction of space around a given point occupied by empty sites ($\phi_0$) and by individuals of the species $i$ ($\phi_i$), satisfying the constraint $\phi_0+\phi_1+ \ldots +\phi_{N}=1$. The mean field equations of motion
\begin{eqnarray}
{\dot \phi}_{0} &&= D \nabla^2 \phi_0 -
r \phi_0 \sum_{i=1}^{N}\phi_i+p\, \sum_{i=1}^{N}\phi_{i}\phi_{i+1}\ ,
\label{1}\\
{\dot \phi}_{i} &&= D \nabla^2 \phi_i +r \phi_0 \phi_i -p\,\phi_{i} \phi_{i-1},\label{2}
\end{eqnarray}
describe the average dynamics of the models studied in the previous Section. In the above equations, a dot stands for the time derivative, $\nabla^2$ is the Laplacian and $D$ is the diffusion rate.


We performed a set of two-dimensional mean field theory simulations starting with initial conditions satisfying $\phi_i=1$ if $i=j$, $\phi_i=0$ if $i \neq j$, where a species $j$ was randomly selected with uniform probability at every grid point ($\phi_0$ was initially set to zero at every grid point). Snapshots of two-dimensional $512^2$ numerical mean field  simulations (using $D=0.5$, $r=0.25$ and $p=0.5$) taken after 5000 generations for $N=4, 6, 8,$ and $10$ are shown in Fig. \ref{fig3}. The results provided by the mean field simulations are consistent with those obtained from the stochastic network simulations discussed in the previous section. While for $N=4$ and $N=6$ (top left and right panels, respectively), no stable dynamical structures develop along the interfaces, it is clear that they do form in the $N=8$ and $N=10$ cases (bottom left and right panels, respectively). As expected, the field theory simulations also show that for $N=10$ the internal structures are more complex than in the $N=8$ case.


\begin{figure}
\centering
\includegraphics[scale= 0.65]{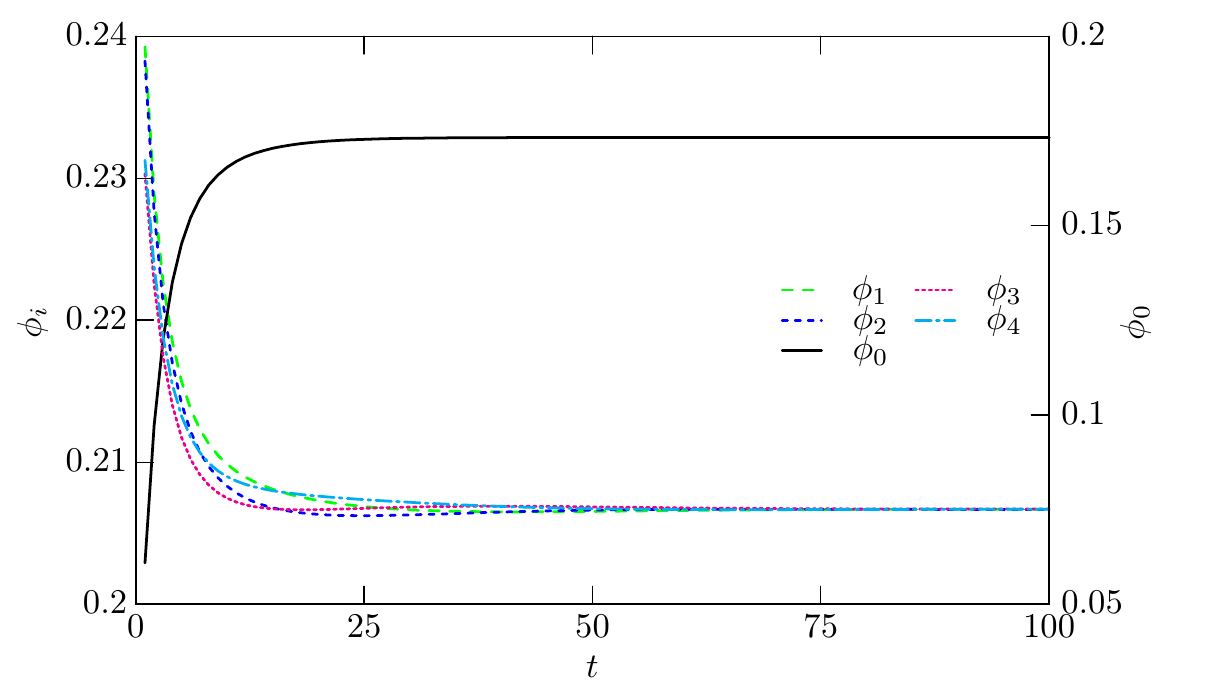}
\caption{(Colour online) Time evolution of $\phi_0$ (solid black line) and $\phi_i$ (coloured lines) at the interface for the model with $4$ species. No prominent dynamical structures are formed at the interfaces in this case.}\label{fig4}
\end{figure}

\begin{figure}
\centering
\includegraphics[scale= 0.65]{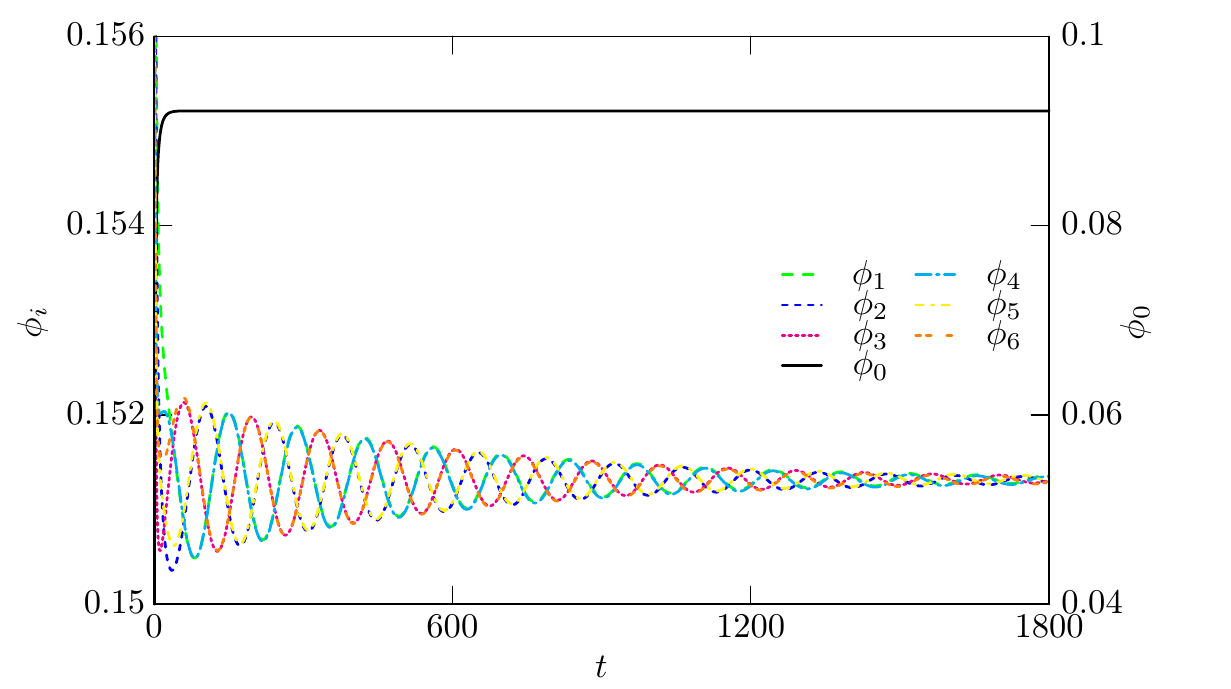}
\caption{(Colour online) Time evolution of $\phi_0$ (solid black line) and $\phi_i$ (coloured lines) at the interface for the model with $6$ species. Transient dynamical structures are formed at the interfaces but they are rapidly damped.} \label{fig5}
\end{figure}

\section{Stability of the dynamical structures at the interfaces}

In order to better resolve the evolution of the dynamical structures at the interfaces, we perform one-dimensional mean field theory simulations for the models described above. Here, we set initial conditions, where the left/right domains of the grid are homogeneously populated with $\phi_i=2/N$, for odd/even $i$ (with $\phi_i$=0 for even/odd $i$ and $\phi_0=0$), respectively. At the interface, the site located at the position ${\cal N}/2$ was initially populated with $\phi_1=1$ (and $\phi_0=\phi_i=0$ for all $i \neq 1$) while the site located at the position ${\cal N}/2+1$ was populated with $\phi_2=1$ (and $\phi_0=\phi_i=0$ for all $i \neq 2$). We verified that our main results are not strongly dependent on the particular choice of initial conditions at the interface. In these simulations we consider $r=0.25$ and $p=0.5$, as before. Given that the thickness of the interfaces is proportional to $D^{1/2}$ here we choose a larger value of $D$ ($D=250$) in order to better resolve the dynamics of the interfaces.

\begin{figure}
\centering
\includegraphics[scale= 0.65]{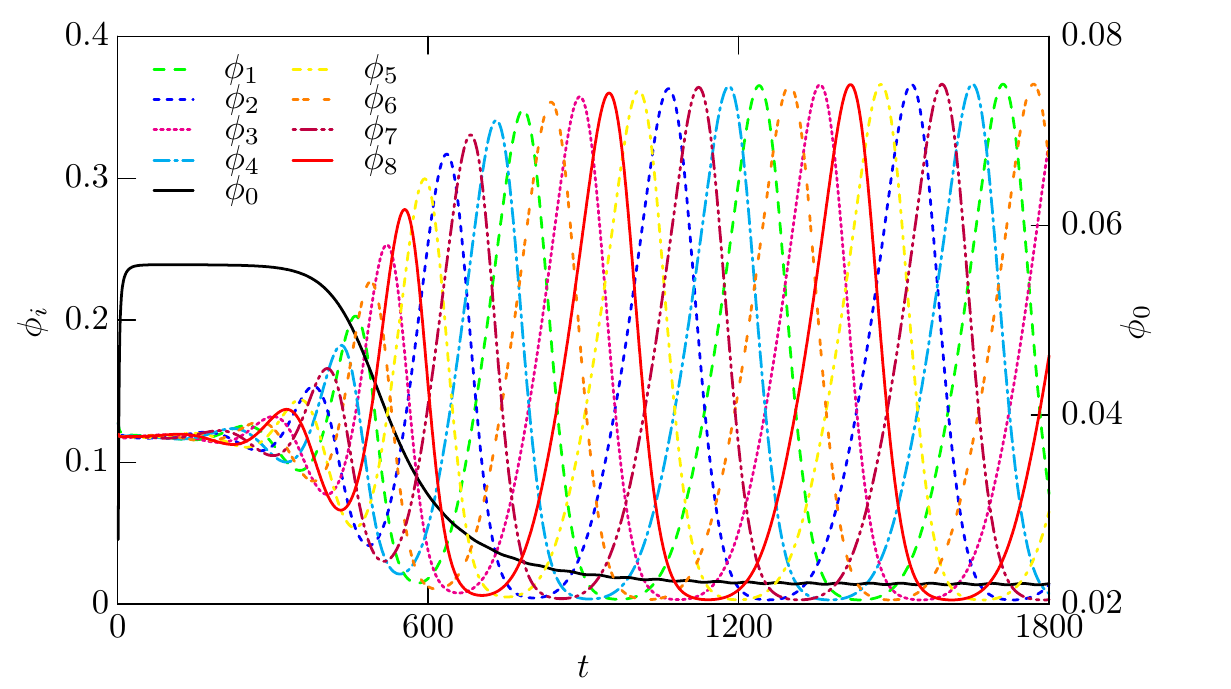}
\caption{(Colour online) Time evolution of $\phi_0$ (solid black line) and $\phi_i$ (coloured lines) at the interface for the model with $8$ species. The oscillation amplitudes of $\phi_i$ increase until reaching a constant value, indicating the stability of the dynamical structures at the interface.} \label{fig6}
\end{figure}

\begin{figure}
\centering
\includegraphics[scale= 0.65]{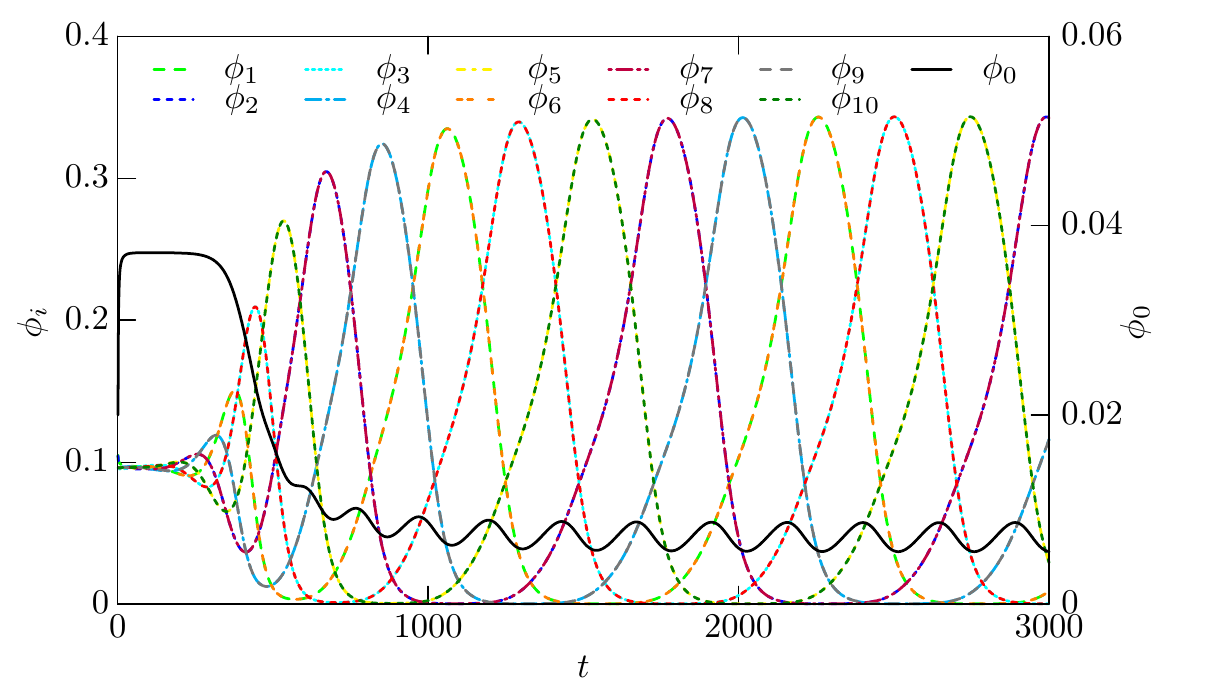}
\caption{(Colour online) Time evolution of $\phi_0$ (solid black line) and $\phi_i$ (coloured lines) for the model with $10$ species. Similarly to the $N=8$ case, the results indicate the stability of the dynamical structures at the interface. However, the oscillatory behaviour of $\phi_0$ is much more pronounced for $N=10$ than for $N=8$.} \label{fig7}
\end{figure}

We start investigating the model with $N=4$, even though it does not support stable dynamics structures at the interfaces. Fig.~\ref{fig4} shows the time evolution of $\phi_0$ (solid black line) and $\phi_i$ (coloured lines) at the interface for the model with $4$ species. We observe no prominent dynamical structures in this case, with the values of $\phi_i$ and $\phi_0$ rapidly approaching their constant asymptotic values. 

A slightly different behaviour can be observed for $N=6$, as we show in Fig.~\ref{fig5}. For $N=6$ there are three couples of mutually neutral individuals belonging to enemy partnerships ($\{1,4\}, \{2,5\}, \{3,6\}$). This is responsible for the transient dynamical structures appearing in Fig.~\ref{fig5}, which nevertheless are rapidly damped. Similarly to the $N=4$ case, at late times the values of $\phi_i$ and $\phi_0$ approaching their constant asymptotic values and no prominent dynamical structures survive at the interface.

On the other hand, Fig. \ref{fig6} shows that for $N=8$ stable dynamical structures develop at the interface. It is possible to observe in Fig. \ref{fig6} the periodic process of creation and annihilation of couples of mutually neutral individuals belonging to enemy partnerships. The values of $\phi_i$ change in time, evolving to become periodic, with constant amplitude, leading to  stable dynamical structures at the interface. In this model there are eight possible couples of mutually neutral individuals belonging to enemy partnerships $({i, i \pm 3})$. These pairs are continuously created and destroyed, resulting in a specific sequence of minimums and maximums $(\{...,1, 4, 7, 2, 5, 8, 3, 6, 1, ...\})$ in Fig.~\ref{fig6}. Note that in general there is always one dominant species belonging to one of the partnerships which is associated to an asymmetric evolution of the interface profile.
 
Finally Fig.~\ref{fig7} shows the time evolution of $\phi_0$ and $\phi_i$ for the model with $10$ species. In this model there are $15$ possible pairs of mutually neutral species belonging to enemy partnerships, $({i, i \pm 3})$ and $({i, i \pm 5})$ (each species can form a couple with three different ones from the enemy partnership). This is responsible for the double sequence of minimums and maximums ($\{...;(1, 6); (8,3); (5,10); (2,7); (4,9);...\}$) observed in Fig.~\ref{fig7}. In this case there is never a dominant partnership, which is responsible for a symmetric evolution of the interface profile. This happens whenever $N/2$ is odd, since in that case for $N \neq 2$, the species $i$ and $i+N/2$ are mutually neutral and belong to enemy partnerships. Both for $N=8$ and $N=10$ the values of $\phi_i$ oscillate periodically, ensuring stability of the dynamical structures formed at the interface. Note, also that the oscillatory behaviour of $\phi_0$ is much more pronounced for $N=10$ than for $N=8$.

\begin{figure}
\centering
\includegraphics[scale=0.64]{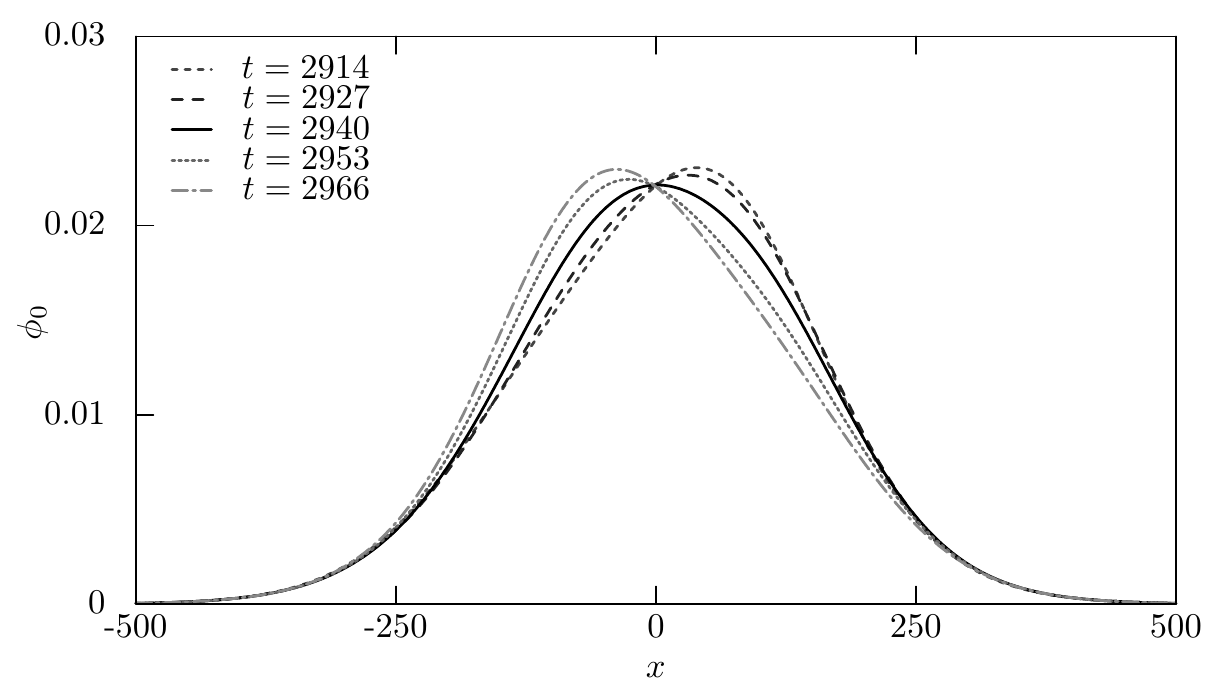}
\includegraphics[scale=0.65]{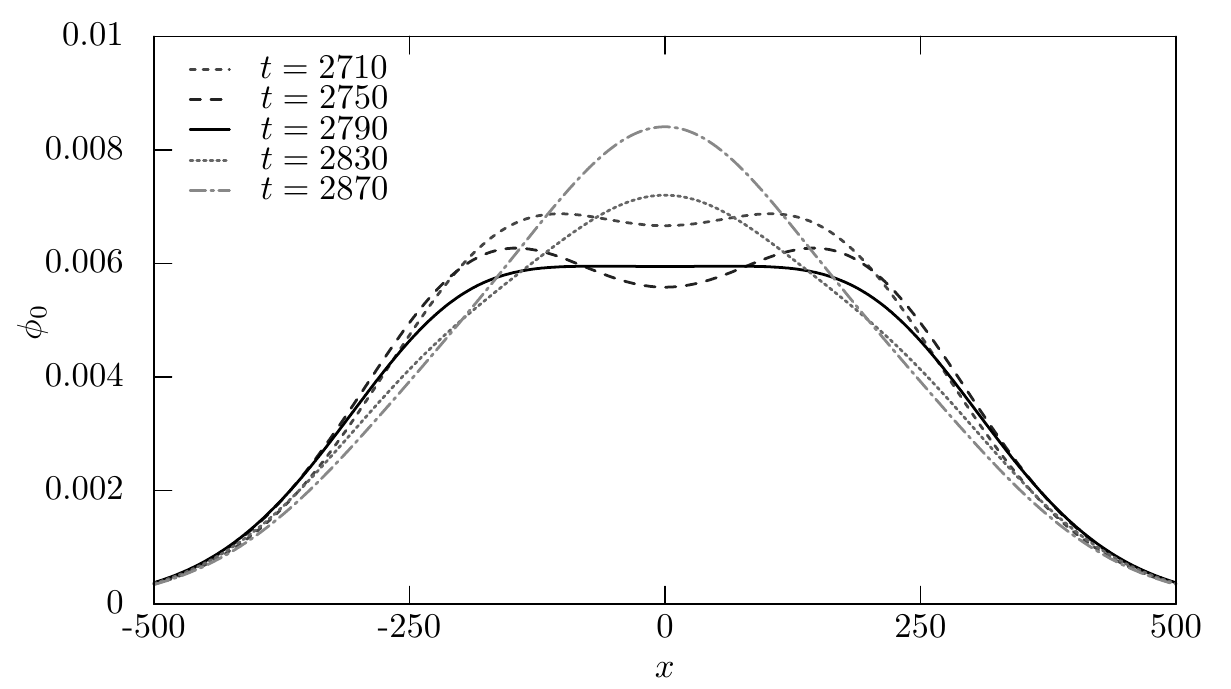}
\includegraphics[scale=0.64]{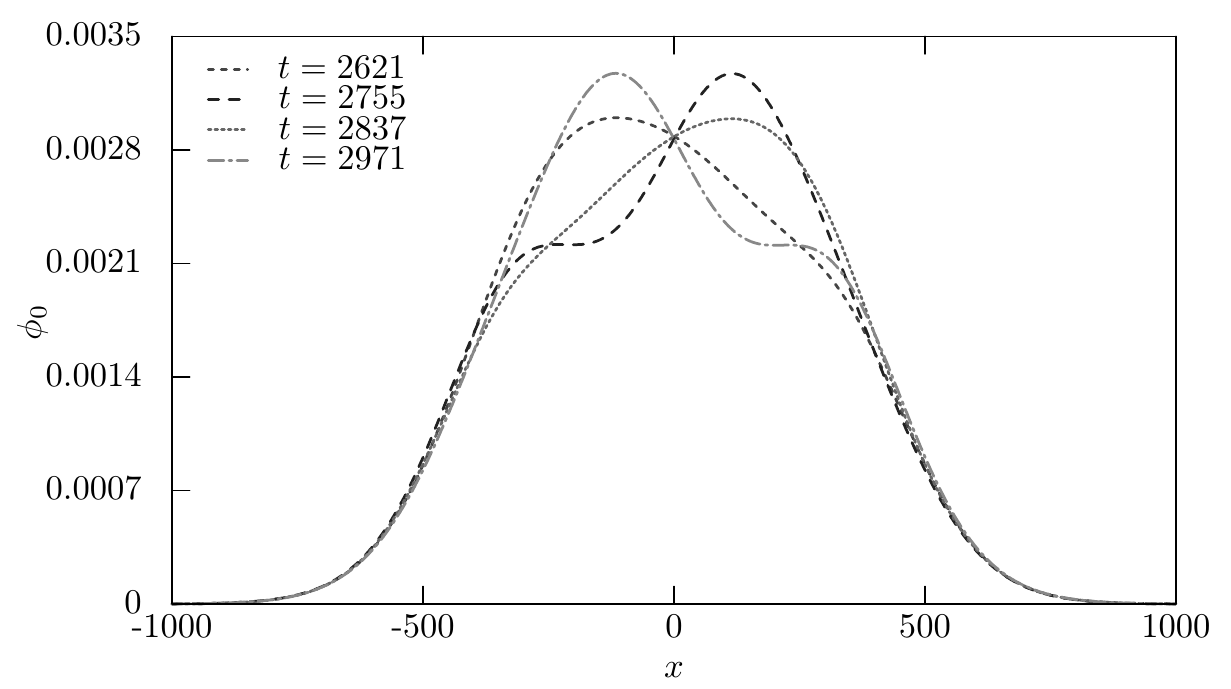}
\includegraphics[scale=0.65]{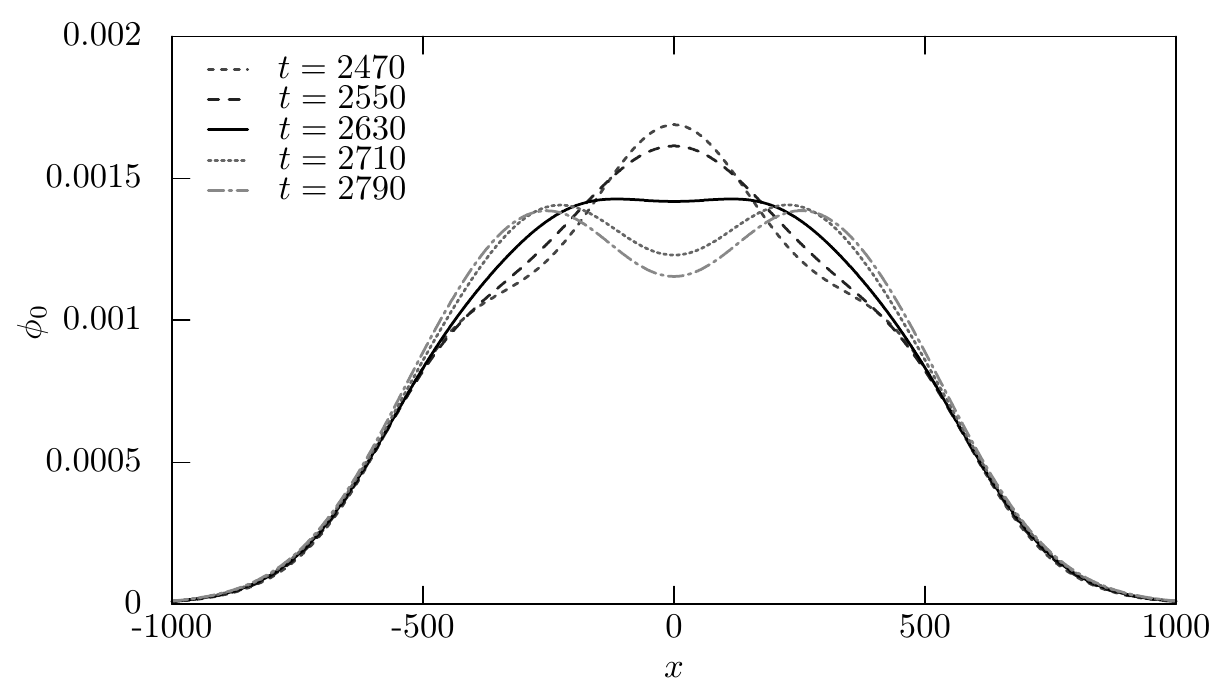}
\caption{Interface profiles $\phi_0(x)$ for different values of the time $t$, showing a symmetric or asymmetric evolution, depending on whether $N/2$ is odd (second (N=10) and fourth (N=14) panels) or even (first (N=8) and third (N=12) panels), respectively.} \label{fig8}
\end{figure}

Fig.~\ref{fig8} shows the evolution of the interface profiles $\phi_0(x)$ for different values of the time $t$. They show a symmetric or asymmetric evolution, depending on whether $N/2$ is odd (second (N=10) and fourth (N=14) panels) or even (first (N=8) and third (N=12) panels), respectively, thus confirming the behaviour discussed above. The movies in  Ref. \cite{video1} and \cite{video2} illustrate the dynamical behaviour of the different species at the interface for $N=8$ (asymmetric evolution of $\phi_0(x)$) and $N=10$ (symmetric evolution of $\phi_0(x)$).

\begin{figure}
\centering
\includegraphics[scale=0.94]{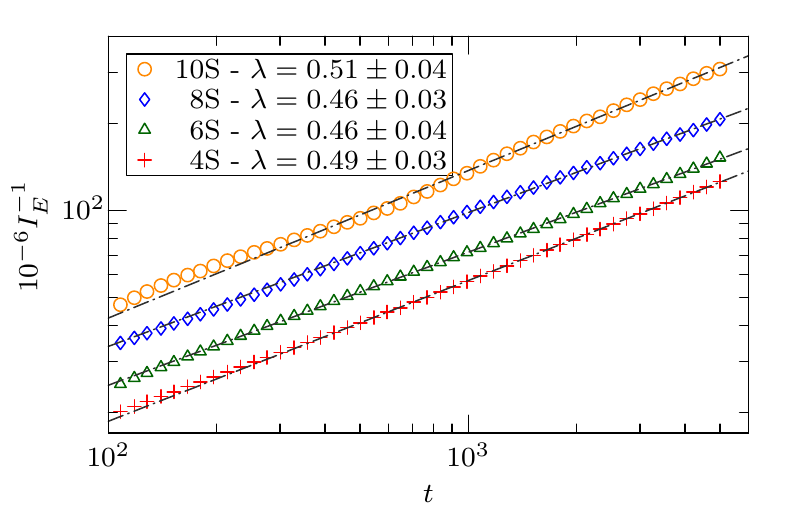}
\includegraphics[scale=0.94]{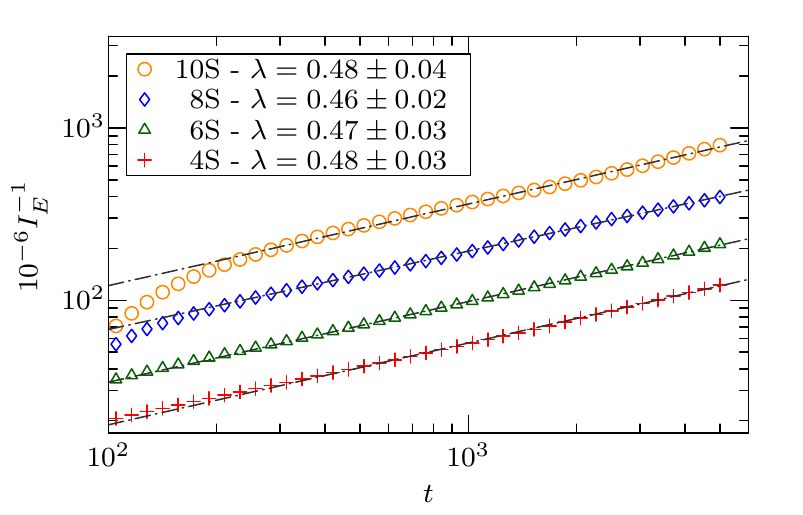}
\caption{(Colour online) Late time behaviour of the length scale $L$, computed from ensembles of twenty $1024^2$
stochastic network (top panel) and mean field theory (bottom panel) simulations, for models with $N=4$, $6$, $8$ and $10$ species.}\label{fig9}
\end{figure}

\section{Macroscopic behaviour of interface networks}


Let us now focus on the macroscopic evolution of the interface networks in order to determine whether or not the presence of stable dynamical structures along the interfaces might leave an imprint on the macroscopic dynamics of the network. The characteristic length of the network can be defined as $L \equiv A/L_{T}$, where $A$ is the (constant) area of the simulation box (proportional to the total number of sites of the grid $\mathcal{N}$) and $L_{T}$ is the total length of the interfaces. Taking into account that the average width and profile of the interfaces remains fixed throughout the simulations, the number of empty sites per unit length is approximately constant and consequently $L_T$ is roughly proportional to the total number of empty sites $I_E$. This implies that the length scale $L$ is inversely proportional to the number of empty sites, that is,
\begin{equation}
L \propto 1/{I_E}\,,
\end{equation}
which follows from \cite{PhysRevE.86.036112,PhysRevE.86.031119}.

The average evolution of $L \propto I_E^{-1}$ with time $t$ was calculated by carrying out sets of 25 distinct two-dimensional stochastic network and mean field theory simulations with distinct random initial conditions. We found that the scaling law $L\propto t^{\lambda}$ describes well the late time evolution of the interface networks investigated in the present paper.

The top panel of Fig.~\ref{fig9} shows the results for the evolution of the characteristic scale $L$ with time $t$ obtained stochastic network simulations, for different values of $N$. We found that $\lambda=0.49 \pm 0.03$, $\lambda=0.46 \pm 0.04$, $\lambda=0.46 \pm 0.03$, $\lambda=0.51 \pm 0.04$, for $N=4$, $6$, $8$ and $10$, respectively. In Fig.~\ref{fig9}, the bottom panel shows analogous results, using mean field theory simulations. The results of the stochastic and mean field network simulations agree well, indicating that the evolution of the characteristic scale $L$ of the network is not significantly affect by the presence of dynamical structures along the interfaces. In all the cases we find a result consistent with the scaling coefficient expected in the case of curvature driven dynamics ($\lambda = 1/2$).


\section{Ending comments}

In this work we described the development of dynamical structures along the interfaces separating domains belonging to enemy partnerships. This was done in the context of cyclic predator-prey models with an even number of species which naturally lead to the partition of the $N$ species into two distinct partnerships. We have shown that as the number of species $N$ increases the number of peaceful associations at the interfaces of mutually neutral individuals also increases, inducing the generation of stable dynamical structures at the interfaces whose complexity also increases with $N$. This behaviour was confirmed using both stochastic and field theory simulations in one and two spatial dimensions. We have also shown that the evolution of the interface profiles can be symmetric or asymmetric depending on whether $N/2$ is odd or even, respectively. We have illustrated this behaviour in several figures throughout the paper as well as in the movies \cite{video1,video2}. Finally we have shown that the internal structure details at the interfaces does not appear to produce any significant changes with respect to the standard macroscopic dynamical evolution of curvature driven interface networks, with the scaling exponent $\lambda$ being consistent with the standard value $\lambda=1/2$ in all cases studied in this paper.

The dynamics of boundary layers is also discussed in detail in \cite{PhysRevE.76.051921}. However, their numerical results are significantly different from ours since in their model no empty site is created when a prey is hunted. The authors stressed the formation of constellations of species in the boundary layers between the alliances. However these patterns arise for a narrow range of mobility rate and the species organise themselves such that predators and preys compose adjacent domains. In contrast, in our model the structures are generated at the interfaces for any choice of the diffusion parameter and they are formed by pairs of mutually neutral species. 

Although, the present work focused on cyclic predator prey-models with an even number of species, many of the results are expected to be quite generic in the context of spatial stochastic May-Leonard models. Once we consider more general classes of models, the presence of structures at the interfaces is no longer restricted to models with even $N$. In fact, if one considers a model with $N$ species where the predation ($ i\  (i+\alpha) \to i\ \otimes$) probabilities $p$ are non-zero for $\alpha=1,...,n-1$ (and zero for other values of $\alpha$), the species separate themselves in $n$ alliances of $N/n$ species. In this case the domains are separated by interfaces with stable dynamical structures if $N/n \geq 4$. For example, for $N=12$ if the species are separated in two or three alliances, stable dynamical structures may develop at the interfaces. Although many these models lead to more complex network patterns with Y-type  and higher order junctions \cite{PhysRevE.86.036112,PhysRevE.86.031119}, the dynamics of the structures at the interfaces is essentially analogous to that studied in the present paper in the context of simpler models. The generalisation of the analysis to models whose food webs are more complex and involve less mutually neutral pairs as in \cite{Szabo2008,1742-5468-2012-07-P07014,PhysRevE.86.036112,PhysRevE.86.031119,PhysRevE.64.042902,0305-4470-38-30-005,Lutz2013286} shall be left for future work.

\begin{acknowledgments}

We thank FCT-Portugal, FAPESP, CAPES/Nanobiotec, CNPq and Fapern for
financial support. The work of P.P.A. was supported by Funda\c c\~ao para a Ci\^encia e a Tecnologia
(FCT) through the Investigador FCT contract of reference IF/00863/2012 
and POPH/FSE (EC) by FEDER funding through the program "Programa
Operacional de Factores de Competitividade - COMPETE.

\end{acknowledgments}


\bibliography{ABLMO3001}

\end{document}